# Edge effects on the electronic properties of phosphorene nanoribbons

Xihong Peng,[1*] Qun Wei,[1,2] Andrew Copple[3]

[1]School of Letters and Sciences, Arizona State University, Mesa, Arizona 85212, USA

[2]School of Physics and Optoelectronic Engineering, Xidian University, Xi'an, 710071, P.R. China

[3]Department of Physics, Arizona State University, Tempe, Arizona 85287, USA

## ABSTRACT

Two dimensional few-layer black phosphorus crystal structures have recently fabricated and demonstrated great potential in applications of electronics. In this work, we employed first principles density functional theory calculations to study the edge effects and quantum confinement on the electronic properties of the phosphorene nanoribbons (PNR). Different edge functionalization groups, such as H, F, Cl, OH, O, S, and Se in addition to a pristine case, were studied for a series width of the ribbon up to 3.5 nm. It was found that the armchair-PNRs (APNRs) are semiconductors for all edge groups considered in this work. However, the zigzag-PNRs (ZPNRs) show either semiconductor or metallic behavior in dependence on their edge chemical groups. Family I edges (H, F, Cl, OH) form saturated bonds with P atoms and the edge states keep far away from the band gap. However, Family II edges (pristine, O, S, Se) form weak unsaturated bonds with the $p_z$ orbital of P atoms and bring edge states within the band gap. These edge states of Family II ribbons present around the Fermi level within the band gap, which close up the band gap of the ZPNRs. For the APNRs, these edge states are at the bottom of the conduction band and result in a reduced band gap.

**Keywords:** phosphorene nanoribbons, edge functionalization group, edge states, band structure, band gap, quantum confinement.

---

[*] Author to whom correspondence should be addressed. Electronic mail: Xihong.peng@asu.edu.

## 1. Introduction

Recently fabricated two dimension (2D) few-layer black phosphorus [1-4] has drawn immediately attention to the society of material science. [5-16] It was found this material is chemically inert and has great transport properties. It was reported that it has carrier mobility up to 1000 $cm^2/V·s$ and an on/off ratio up to $10^4$ was achieved for the phosphorene transistors at room temperature. [2, 17] In addition, this material shows a finite direct band gap at the Γ point of Brillouin zone [2, 5, 7, 17-19] (in contrast to the vanishing gap in graphene), which opens doors for additional applications in optoelectronics.

Tailoring electronic properties of semiconductor has been critical for applications in electronics. A series of strategies were explored to engineer the band gap of phosphorene, for example, by utilizing multilayer structures, [5, 17] applying mechanical strains, [6, 7] forming nanoribbons [9, 12-14] or nanotubes [9]. For the phosphorene nanoribbons (PNRs), their electronic properties are dependent on the crystal orientation of the ribbons. For example, two typical crystal directions were generally explored, namely the armchair-PNRs (APNRs) and the zigzag-PNRs (ZPNRs). Tran and Yang [12] reported that the PNRs with the edge P atoms passivated using H are direct-gap semiconductors and their band gaps are a strong function of the ribbon width due to quantum confinement effect. However, Guo et. al. [9] found that the pristine ZPNRs are metals regardless of the ribbon width, while the pristine APNRs are semiconductors with indirect band gaps. These distinct conclusions imply that the edges of the ribbons play a critical role on their electronic properties. Therefore, it is of importance to systematically study the edge effects on the PNRs, in particular, with several common chemical groups, such as -OH, -O, -S etc. In this work, we present detailed systematic analysis of the edge effects on the electronic band structure and density of states (DOS) of both APNRs and ZPNRs for a series of widths up to 3.5 nm. Our results suggest that the APNRs are semiconductors with either direct or indirect band gap depending on the edge function groups, and the ZPNRs demonstrate either semiconductor or metallic behavior with different edge passivation.

## 2. Simulation details

The theoretical calculations were carried out using first principles density functional theory (DFT).[20] The Perdew-Burke-Ernzerhof (PBE) exchange-correlation functional [21] and the projector-augmented wave (PAW) potentials [22, 23] were employed. The calculations were performed using the Vienna Ab-initio Simulation Package (VASP).[24, 25] The kinetic energy cutoff for the plane wave basis set was chosen to be 500 eV. The energy convergence criteria for electronic and ionic iterations were set to be $10^{-5}$ eV and $10^{-4}$ eV, respectively. The reciprocal space was meshed at $14 \times 1 \times 1$ for the ZPNRs and $1 \times 10 \times 1$ for the APNRs using Monkhorst Pack meshes centered at Γ point. 21 K-points were included in band structure calculations from Γ to X for the ZPNRs and from Γ to Y for the APNRs. To simulate a ribbon, a unit cell with periodic boundary condition was used. A vacuum space of at least 20 Å was included in the unit cell to minimize the interaction between the system and its replicas resulting from the periodic boundary condition.

## 3. Results and discussion

### A. Structural properties

The initial structures of monolayer phosphorene were obtained from bulk black phosphorus.[26] The 2D phosphorene has a puckered honeycomb structure with each phosphorus atom covalently bonded with three adjacent atoms. Our calculated lattice constants for bulk black phosphorus are $a = 3.307$ Å, $b = 4.547$ Å, and $c = 11.210$ Å, in good agreement with experimental values [26] and other theoretical calculations.[17, 27] The relaxed lattice constants for monolayer phosphorene are $a = 3.295$ Å, $b = 4.618$ Å.

The APNRs and ZPNRs with different ribbon widths up to 3.5 nm were truncated from monolayer phosphorene along the y- and x-directions, respectively, as shown in Figure 1. The width of a ribbon $n$L is referred according to the number $n$ of P atoms in the direction perpendicular to the ribbon direction (see Figure 1). As an example, Figure 1 demonstrates the snapshots of 9L-APNRs and 12L-ZPNRs. The edges of the PNRs were treated in eight different scenarios: no passivation (pristine) or bonded with H, F, Cl, OH, O, S or Se chemical species.

We explored the structural configuration at the edges for the PNRs. For example, six bond lengths labeled as b1 – b6 and four bond angles indicated as α, β, γ and θ in Figure 1 were calculated and reported in Table I. The bond length b1 between the two P atoms near the edge of the 9L-APNR has a negligible change for all eight different edge groups, indicating that the distinct edge function groups only affect the geometry of the ribbon at the edge. The P-P bond b2 at the edge for the pristine case experience a considerable reduction from 2.26 Å (of the 2D phosphorene) to 2.07 Å, due to its edge dangling-bond reconstruction. The variation in the bond length b3 between the P atoms and the edge species is expected: larger edge chemical species has a longer bond length. The bond b3 is sufficient large for the edge S (Se) case, so that two S (Se) atoms in the neighbored simulation cell form a bond, labeled as b4 in Figure 1(a). The bond angles α and β both increase largely for the pristine case due to the edge P-P bond reconstruction.

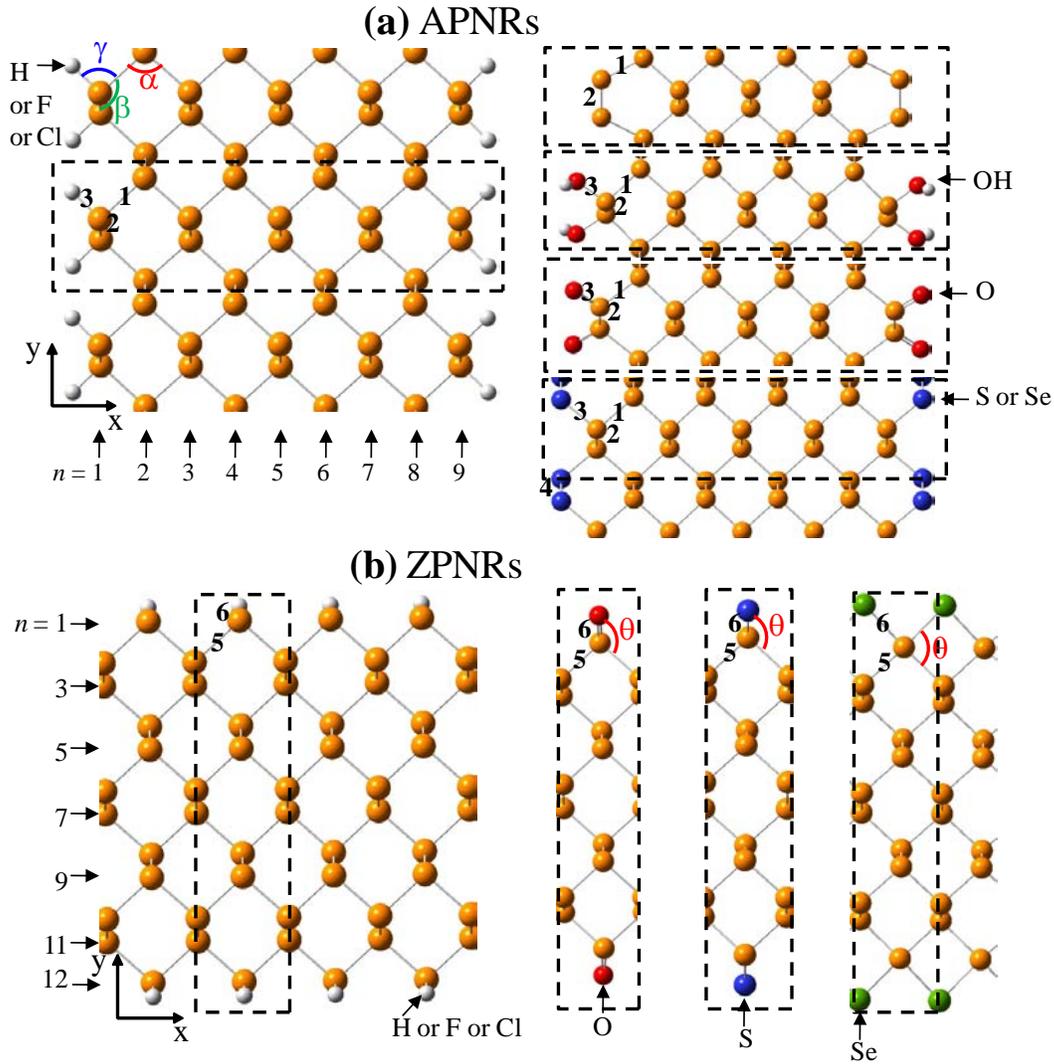

**Figure 1. Snapshots of the APNRs and ZPNRs with different edge functionalization groups. (a) 9L-APNR with edge P atoms saturated using H (F or Cl), and hydroxyl group, double-bonded O, and bridge-bonded S (Se) atoms, respectively. (b) 12L-ZPNR with edge functionalized using H (F or Cl), double-bonded O, S and bridge-bonded Se atoms, respectively. The dashed rectangles indicate the unit cells.**

**Table I,** The bond lengths b1-b6 and bond angles α-θ at the edges of the PNRs with different edge groups. The bond lengths and angles were denoted in Figure 1. As a reference, the corresponding bond lengths/angles in monolayer phosphorene were also listed.

| system | 9L-APNR | | | | | | | 12L-ZPNR | | |
|---|---|---|---|---|---|---|---|---|---|---|
| | b1 (Å) | b2 (Å) | b3 (Å) | b4 (Å) | α (°) | β (°) | γ (°) | b5 (Å) | b6 (Å) | θ (°) |
| monolayer | 2.22 | 2.26 | n/a | n/a | 95.9 | 104.1 | n/a | 2.22 | n/a | 104.1 |
| pristine | 2.23 | 2.07 | n/a | n/a | 111.1 | 119.1 | n/a | 2.14 | n/a | n/a |
| H | 2.22 | 2.25 | 1.44 | n/a | 95.7 | 103.2 | 93.1 | 2.23 | 1.44 | 99.3 |
| F | 2.23 | 2.25 | 1.63 | n/a | 94.9 | 98.3 | 98.3 | 2.22 | 1.64 | 106.0 |
| Cl | 2.24 | 2.26 | 2.08 | n/a | 92.7 | 95.9 | 101.1 | 2.23 | 2.08 | 107.3 |
| OH | 2.23 | 2.24 | 1.68 | n/a | 94.9 | 99.0 | 100.1 | 2.22 | 1.68 | 110.7 |
| O | 2.27 | 2.25 | 1.49 | n/a | 99.5 | 106.7 | 109.2 | 2.25 | 1.50 | 116.4 |
| S | 2.23 | 2.25 | 2.11 | 2.10 | 97.0 | 105.0 | 99.5 | 2.28 | 2.00 | 106.3 |
| Se | 2.23 | 2.24 | 2.28 | 2.38 | 95.3 | 102.9 | 100.0 | 2.29 | 2.39 | 89.3 |

For the ZPNRs, the bond length b5 and b6 show a similar variation with the edge function groups as the APNRs. The significantly reduced bond angle θ for the Se case from 104.1° (of the 2D phosphorene) to 89.3° is resulted from the special bridge-bonded configuration as shown in Figure 1(b). We also checked this bridge-bonded arrangement for both O and S edges and found that these two prefer to binding to one P atom instead of bridge-bond to two P atoms as shown in Figure 1(b). This is resulted from their much shorter bond lengths b6 with P, 1.50 Å and 2.00 Å for the O and S cases, respectively, compared to 2.39 Å in the Se edge.

### B. Band structure and density of states

The band structures of the PNRs with the eight different edge functionlization groups were calculated. As an example, Figure 2 presents the band structure of the 9L-APNRs and 12L-ZPNRs. Since the edge groups F and Cl show similar effects on the band structure, we only plot the F-edge case for the APNR and the Cl case for the ZPNR in Figure 2. It is clear that the APNR is a semiconductor. The band gap is defined as the energy difference between the conduction band minimum (CMB) and valence band maximum (VBM). For the pristine and O-edge cases, the APNR shows an indirect band gap, while other function groups demonstrate a direct band gap. For the pristine and O-edge cases, the CBM, which is contributed from the edge P and O atoms, respectively (see below Figure 4), is not located at the Γ point. While the VBM is at Γ, which gives an indirect gap. However, for the cases with the edge H, OH, F(Cl), both the CBM and VBM are contributed by the non-edge P atoms (i. e. intrinsic states) in the ribbon and located at Γ, which gives a direct band gap. A slightly different situation occurs for the APNRs with the edge S (Se) atoms, in which the conduction bands are mainly contributed by the edge S (Se). While the CBM is still located at Γ and results in a direct band gap.

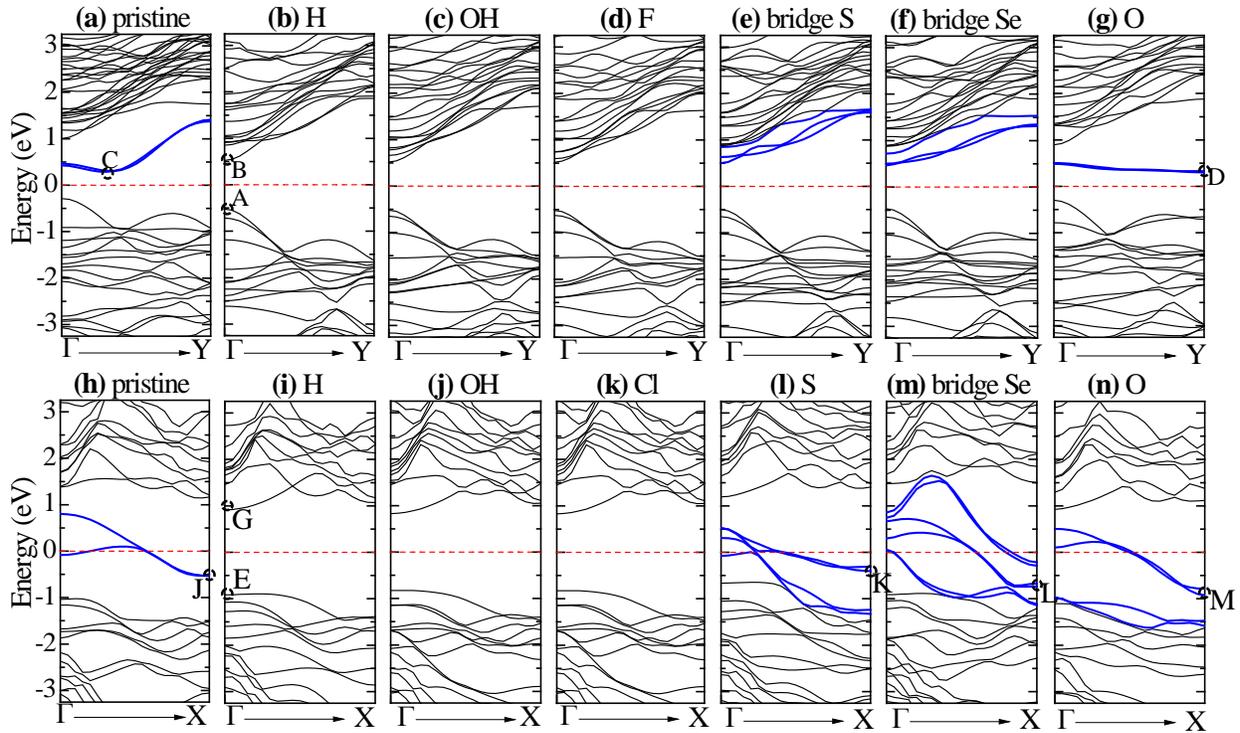

**Figure 2, Band structures of the APNRs and ZPNRs with different edge chemical groups. Top is for the 9L-APNRs and Bottom for the 12L-ZPNR. The Fermi level is set at zero. The edge states brought by the edge P, O, S and Se atoms within the band gap are indicated in blue color.**

On the other hand, the ZPNR shows a dissimilar behavior. The ribbon demonstrates either semiconductor or metallic behavior as shown in Figure 2(h)-(m), in dependence on the edge functionalization groups. The edge chemical groups can be classified into two distinct families. Family I includes the H, OH, F and Cl edges and Family II consists of the pristine, O, S and Se cases. With Family I edges, the ZPNRs are semiconductors with a direct band gap at Γ. And their CBM and VBM are the intrinsic states from the non-edge P atoms in the ribbon (see below Figure 4). However, for Family II edges, the ribbon shows metallic. The electronic states contributed by the edge atoms of Family II are located around the Fermi level and close up the band gap.

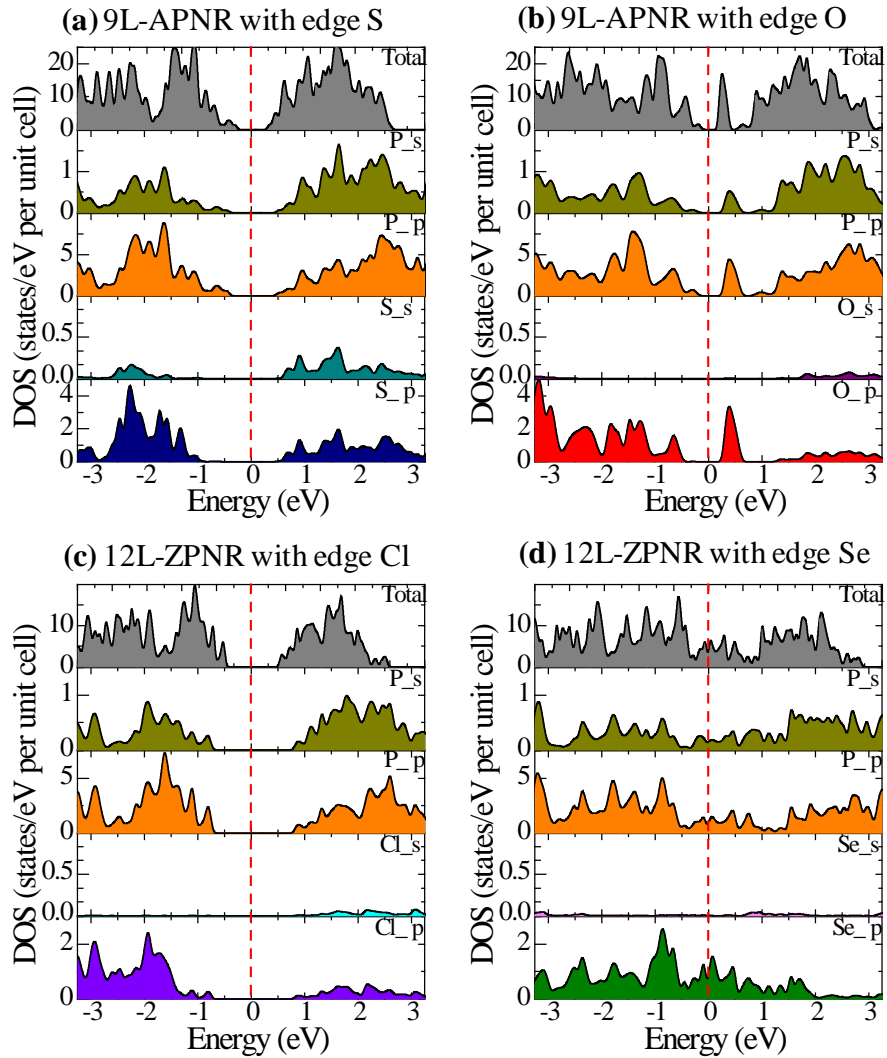

**Figure 3, The total, s- and p-orbital projected density of states. The DOS of the 9L-APNR with the edge functionalized using (a) S and (b) O atoms. The DOS of the 12L-ZPNR with the edge saturated using (c) Cl and (d) bridge-bonded Se atoms. The Fermi level is aligned at zero.**

The DOS of the studied PNRs was also calculated. As an example, Figure 3 presents the total, s- and p-orbital projected DOS of the 9L-APNRs and 12L-ZPNRs. The DOS of the 9L-APNR with the edge S (O) passivation in Figure 3 suggest that the ribbon are semiconductors, in which the conduction band (CB) was mainly contributed by the p-orbitals of the edge P and S (O) atoms while the VBM is located at the p-orbitals of the P atoms in the ribbons. For the 12L-ZPNR with the edge Cl in Figure 3(c), the band gap of the ribbon is determined by the intrinsic states of P and the Cl states are far away from the band gap. In the metallic 12L-ZPNR with the bridge-bonded Se in Figure 3(d), the p-orbitals of the Se and P atoms form bonds which close up the gap.

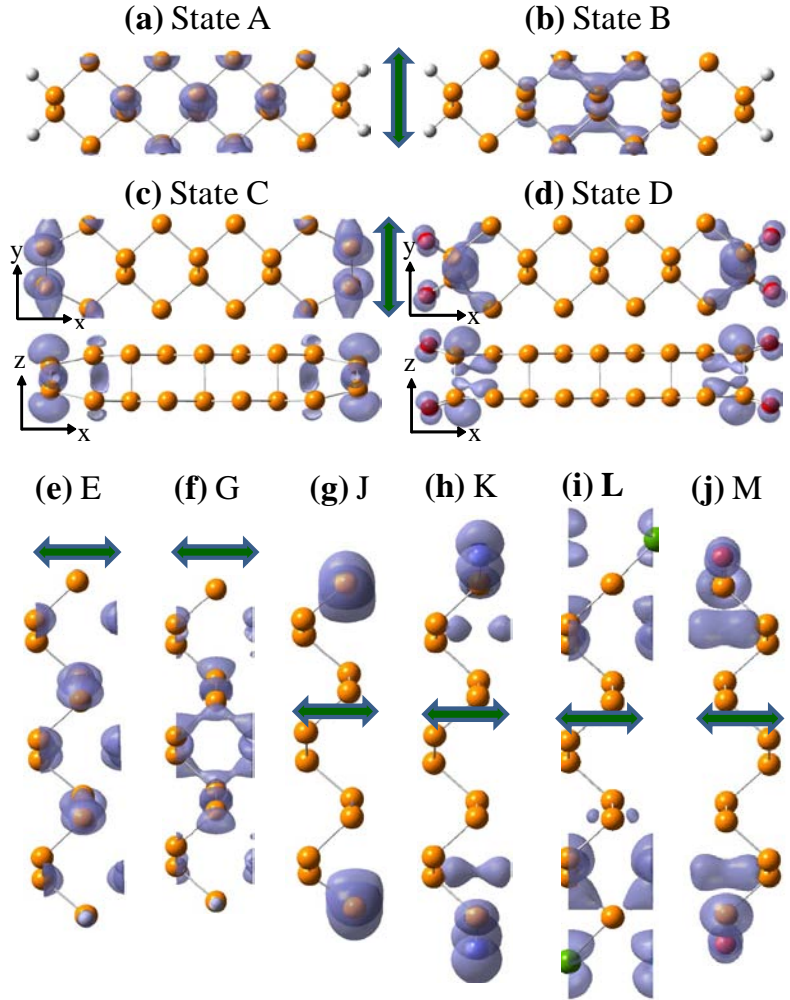

**Figure 4, The electron density contour plots of near-Fermi-level states A – M. The vertical and horizontal arrows indicate the armchair and zigzag ribbon directions, respectively.**

The near-band-edge states A – M (labeled in Figure 2) were explored and their electron density contour plots are presented in Figure 4. States A and B are the VBM and CBM of the 9L-APNR with the edge functionalization groups H, OH, F and Cl (i.e. Family I). The wave function of states A and B are primarily located in the phosphorus atoms within the ribbon and the edge P and chemical groups have little contribution. State C is the CBM of the pristine 9L-APNR. From the electron density contour plot in Figure 4 (c), it is clear that the charge is primarily located at the edge P atoms. State D is the CBM of the 9L-APNR with the edge O and the charge is distributed mainly on the edge P and O atoms. Similar edge states within the band gap were also found for the S and Se cases.

In the 12L-ZPNRs, the electronic states E and G are denoted as the VBM and CBM with the edge H, OH, F and Cl cases (Family I) and they are intrinsic states contributed by the non-edge P atoms. However, for the pristine, S, Se and O cases (Family II), the charges of states J, K, L and M are primarily located on the edge P and chemical groups.

To understand why Family II cases bring edge states within the band gap while Family I edges not, we examined the characteristics of the electronic orbitals of the near-gap states. Family I edge species form a saturated bond with the P atoms in the ribbons and the energy associated with this saturated bond is far below the Fermi level. For example, the energy associated with the P-F bond in the 9L-APNR is 4.12 eV below the Fermi energy and that of the P-O bond in the case of edge OH group is 2.09 eV below the Fermi level. However, in Family II cases, the P atoms don't form a saturated bond with the edge species. Moreover, these unsaturated bonds are particularly weak due to their special electronic orbital orientations. For instance, the edge reconstructed P-P bond in the pristine ribbon is nearly in the ribbon plane (i. e. the xy-plane). However, the p-orbitals of the two P atoms are along the z-direction (i.e. $p_z$-orbital). And the $p_z$-orbitals form a relatively weak P-P bond in the xy-plane due to a minimal overlap of the wavefunction. Similar situation was found for the O, S and Se cases. For example in Figure 4(d), the $p_z$-orbitals of the edge O and P atoms cause a weak P-O bond in the ribbon plane. These weak unsaturated bonds bring the edge states within the band gap for Family II cases. It is interesting to note that the bridge-bonded Se atoms in the ZPNR appear to form a saturated bond with two P atoms. However, the $p_z$-orbitals make these P-Se-P bonds in the ribbon plane are relatively weak, which still bring the edge states within the band gap.

### C. Quantum confinement on band gaps

The band gap of the APNRs and semiconducting ZPNRs with the ribbon width up to 3.5 nm were calculated and presented in Figure 5. For the APNRs in Figure 5(a), the band gap of Family I cases increases rapidly with a reduced width of the ribbon. Note from the above discussion, the band gap of Family I is determined by the intrinsic electronic states of phosphorus atoms. The scaling of the band gap with the ribbon width $d$ obeys the usual $1/d^2$ relation according to quantum confinement, which is consistent with literature. [12] Given a same width, the band gap of Family II ribbons is generally smaller than that of Family I. This is because Family II cases bring edge states within the band gap, thus largely reducing the gap. The significant smaller band gap of the 3L-APNR with the edge O is resulted from a structural distortion of this ultra narrow ribbon. For the ZPNRs in Figure 5(b), the four of Family I ribbons have very similar behavior, and the gap scaled as $1/d$, in agreement with Tran and Yang's prediction. [12] Family II ZPNRs show metallic behavior thus the gap is zero.

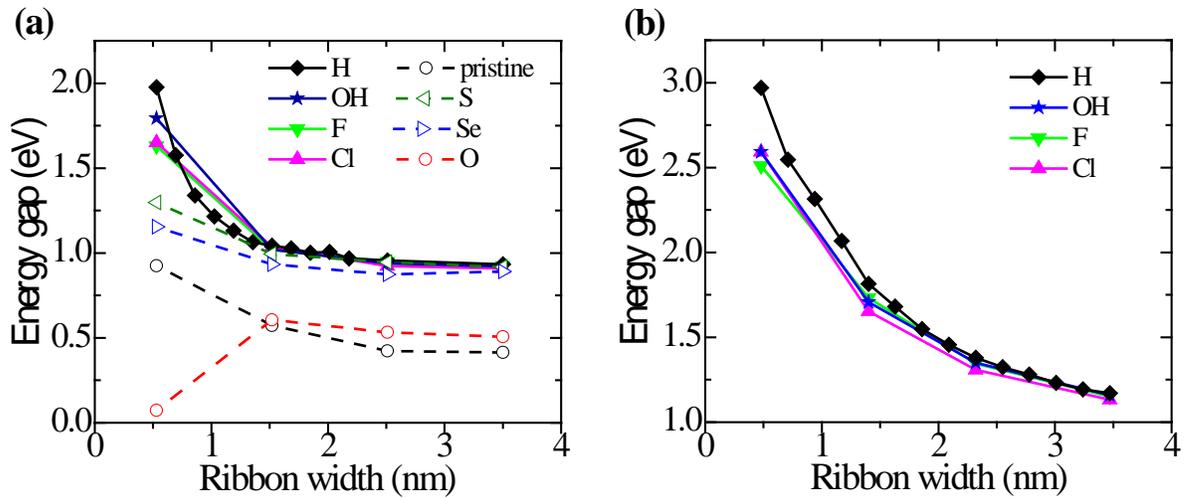

**Figure 5, The band gap of the (a) APNRs and (b) ZPNRs as a function of ribbon width. Family I edges are plotted using solid lines while Family II edges denoted with dashed lines.**

## 4. Conclusion

We employed first principles DFT calculations to study electronic properties of the phosphorene nanoribbons with the edge functionalized using different chemical groups. It was found that the APNRs are semiconductors for all edge groups considered in this work. However, the ZPNRs demonstrate either semiconducting or metallic behavior. The CBM and VBM of the APNRs and ZPNRs with the edge H, F, Cl, and OH groups (Family I edges) are contributed by the intrinsic electronic states of non-edge phosphorus atoms, and the edge species have negligible contribution to their wavefunctions. Therefore, the ribbons in this family are semiconductors with a direct band gap. However, the APNRs and ZPNRs in the pristine, O, S and Se cases (Family II edges) display edge states within the band gap, which cause a reduced band gap in the APNRs and metallic behavior in ZPNRs. These edge states in Family II ribbons are originated from their weak unsaturated bond with the P atoms.

**Acknowledgement**

This work is supported by the Faculty Research Fund from School of Letters and Sciences at Arizona State University (ASU) to Peng. The authors thank ASU Advanced Computing Center and XSEDE for providing computing resources. Dr. F. Tang is acknowledged for the helpful discussions and critical review of the manuscript.

1. H. Liu, A. T. Neal, Z. Zhu, Z. Luo, X. Xu, D. Tománek, and P. D. Ye, (2014), Phosphorene: An Unexplored 2D Semiconductor with a High Hole Mobility, *Acs Nano* **8**, 4033-4041.
2. Likai Li, Yijun Yu, Guo Jun Ye, Qingqin Ge, Xuedong Ou, Hua Wu, Donglai Feng, Xian Hui Chen, and Yuanbo Zhang, (2014), Black phosphorus field-effect transistors, *arXiv:1401.4117*
3. F. Xia, H. Wang, and Y. and Jia, (2014), Rediscovering Black Phosphorus: A Unique Anisotropic 2D Material for Optoelectronics and Electronics, *arXiv:1402.0270*
4. E. S. Reich, (2014), Phosphorene excites materials scientists, *Nature* **506**, 19.
5. Vy Tran, Ryan Soklaski, Yufeng Liang, and Li Yang, (2014), Layer-Controlled Band Gap and Anisotropic Excitons in Phosphorene and Few-Layer Black Phosphorus, *arXiv:1402.4192*
6. Ruixiang Fei and Li Yang, (2014), Strain-Engineering Anisotropic Electrical Conductance of Phosphorene and Few-Layer Black Phosphorus, *arXiv:1403.1003*
7. X. Peng, A. Copple, and Q. Wei, (2014), Strain engineered direct-indirect band gap transition and its mechanism in 2D phosphorene, *arXiv:1403.3771*
8. J. Dai and X. C. Zeng, (2014), Bilayer Phosphorene: Effect of Stacking Order on Bandgap and its Potential Applications in Thin-Film Solar Cells, *The Journal of Physical Chemistry Letters* **5**, 1289-1293.
9. H. Guo, N. Lu, J. Dai, X. Wu, and X. C. Zeng, (2014), Phosphorene nanoribbons, nanotubes and van der Waals multilayers, *arXiv:1403.6209*
10. Q. Wei and X. Peng, (2014), Superior mechanical flexibility of phosphorene and few-layer black phosphorus, *arXiv:1403.7882*
11. W. Lu, H. Nan, J. Hong, Y. Chen, C. Zhu, Z. Liang, X. Ma, Z. Ni, C. Jin, and Z. Zhang, (2014), Plasma-assisted fabrication of monolayer phosphorene and its Raman characterization, *Nano Research*, DOI:10.1007/s12274-12014-10446-12277.
12. V. Tran and L. Yang, (2014), Unusual Scaling Laws of the Band Gap and Optical Absorption of Phosphorene Nanoribbons, *arXiv:1404.2247*
13. A. Maity, A. Singh, and P. Sen, (2014), Peierls transition and edge reconstruction in phosphorene nanoribbons, *arXiv:1404.2469*
14. A. Carvalho, A. S. Rodin, and A. H. C. Neto, (2014), Phosphorene nanoribbons, *arXiv:1404.5115*
15. H. Y. Lv, W. J. Lu, D. F. Shao, and Y. P. Sun, (2014), Large thermoelectric power factors in black phosphorus and phosphorene, *arXiv:1404.5171*
16. Michele Buscema, Dirk J. Groenendijk, Sofya I. Blanter, Gary A. Steele, Herre S.J. van der Zant, and A. Castellanos-Gomez, (2014), Fast and broadband photoresponse of few-layer black phosphorus field-effect transistors, *arXiv:1403.0565*
17. Han Liu, Adam T. Neal, Zhen Zhu, David Tomanek, and Peide D. Ye, (2014), Phosphorene: A New 2D Material with High Carrier Mobility, *arXiv:1401.4133*
18. Y. Takao and A. Morita, (1981), Electronic structure of black phosphorus: Tight binding approach, *Physica B & C* **105**, 93-98.
19. A. S. Rodin, A. Carvalho, and A. H. Castro Neto, (2014), Strain-induced gap modification in black phosphorus, *arXiv:1401.1801*

20. W. Kohn and L. J. Sham, (1965), Self-consistent equations including exchange and correlation effects, *Physical Review* **140, A1133-A1138.**
21. J. P. Perdew, K. Burke, and M. Ernzerhof, (1996), Generalized Gradient Approximation Made Simple, *Physical Review Letters* **77, 3865-3868.**
22. P. E. Blochl, (1994), Projector augmented-wave method, *Physical Review B* **50, 17953-17979.**
23. G. Kresse and D. Joubert, (1999), From ultrasoft pseudopotentials to the projector augmented-wave method, *Physical Review B* **59, 1758-1775.**
24. G. Kresse and J. Furthmuller, (1996), Efficient iterative schemes for ab initio total-energy calculations using a plane-wave basis set, *Physical Review B* **54, 11169.**
25. G. Kresse and J. Furthmuller, (1996), Efficiency of ab-initio total energy calculations for metals and semiconductors using a plane-wave basis set, *Computational Materials Science* **6, 15-50.**
26. Allan Brown and Stig Rundqvist, (1965), Refinement of the crystal structure of black phosphorus, *Acta Cryst.* **19, 684.**
27. Jingsi Qiao, Xianghua Kong, Zhi-Xin Hu, Feng Yang, and Wei Ji, (2014), Few-layer black phosphorus: emerging 2D semiconductor with high anisotropic carrier mobility and linear dichroism, *arXiv:1401.5045*